# A Magnetization and Microwave Study of Superconducting MgB$_2$

A. Dulcic, D. Paar, M. Požek

*Department of Physics, Faculty of Science, University of Zagreb, P. O. Box 331, HR-10002 Zagreb, Croatia*

G. V. M. Williams

*2. Physikalisches Institut, Universität Stuttgart, D-70550 Stuttgart, Germany and Industrial Research Limited, P.O. Box 31310, Lower Hutt, New Zealand.*

S . Krämer

*2. Physikalisches Institut, Universität Stuttgart, D-70550 Stuttgart, Germany.*

C. U. Jung, Min-Seok Park, Sung-Ik Lee.

*National Creative Research Initiative Center for Superconductivity and Department of Physics, Pohang University of Science and Technology, Pohang 790-784, Republic of Korea*

*10 July, 2001*

**ABSTRACT**

We report magnetic field dependent magnetization and microwave impedance measurements on a MgB$_2$ superconductor prepared by high pressure synthesis. We find that the upper critical field is linearly dependent on temperature near T$_c$ and the dc irreversibility field exponent is ~1.4. The microwave data display an excess surface resistance below T$_c$ which is neither observed in low T$_c$ nor in high temperature superconductors (HTSC). The real part of the complex conductivity, $\sigma_1$, shows a huge maximum below T$_c$ and the imaginary part, $\sigma_2$, is linear for temperatures less than 20 K, which can not be simply accounted for by the weak coupling BCS model with an s-wave superconducting order parameter. We speculate that this may be due to the two gaps reported by other studies. Unlike measurements on the high temperature superconducting cuprates, we find no evidence of weak-links in the superconducting state. By inverting the magnetic field dependent impedance data, we find a vortex depinning frequency that decreases with increasing magnetic field and evidence for an anisotropic upper critical magnetic field.

**Introduction**

The report of superconductivity in commercially available $MgB_2$ is proving to be particularly interesting especially as the superconducting transition temperature is relatively high (39 K) [1]. Recent Hall effect experiments have shown that, like the high temperature superconducting cuprates (HTSC), the carriers in $MgB_2$ are holes [2-4]. The observation of a boron isotope effect has lead to $MgB_2$ being described as a phonon-mediated BCS superconductor [4] although the observed $T_c$ is at, or above, the early theoretical predictions based on phonon-mediated coupling within the BCS theory [6]. It has recently been shown that the $T_c$ can be as large as that observed within the moderate to strong phonon-mediated BCS theory due to the large density of states at the Fermi level and high phonon frequencies [7-9]. However, it has been suggested that the predicted crossing of the Fermi surface by the boron $\sigma$ and $\pi$ bands at different regions of the Brillouin zone can have a significant impact on superconductivity in $MgB_2$ [10,11]. These bands derive from the boron 2D $p_{xy}$ bands in the ab-plane and the isotropic $p_z$ bands which couple the boron planes to each other. It has even been suggested that the electron-phonon mechanism is not important in $MgB_2$ [12].

The understanding of superconductivity in $MgB_2$ is further complicated by conflicting reports concerning the magnitude and symmetry of the superconducting gap. Some studies report a s-wave superconducting order parameter where $2\Delta/k_BT$ is less than that expected within the weak-coupling BCS theory [13-15] or large enough so that $MgB_2$ is in the strong-coupling regime [16]. Other studies report the existence of two superconducting gaps [17-20] or nodes in the superconducting gap [21]. Early $^{11}B$ nuclear magnetic resonance measurements supported the s-wave superconducting order parameter model and a small coherence peak was reported [16]. However, this has been questioned by a later study that accentuated the importance of flux motion to the value of the spin-lattice relaxation rate [22].

There is a large variation in the reported values and temperature dependences of the basic parameters such as the upper critical field, $H_{c2}$, and the magnetic irreversibility field, $H_{irr}$. For example, the anisotropy of $H_{c2}$ is reported to range from 1.23 to 2.6 [23-25] and it has been suggested that the anisotropy could be as high as 9 [26]. There have also been reports that $H_{c2}$ is not linearly dependent on temperature near $T_c$ [23,24,27-29]. It has been suggested that this positive curvature is similar to that observed in the nickel borocarbides and can be explained by a two-band model for superconductivity in the clean limit [27]. However, a linear temperature dependence of $H_{c2}$ was reported in other studies [30,31]. The temperature dependence of $H_{irr}$ has also been investigated [31-34]. For example, two reports of the magnetic irreversibility field have found that $H_{irr}$ scales linearly with the temperature deviation from $T_c$ [31,32] but the analysis of $H_{irr}$ reported in two other studies gives a scaling of the form $H_{irr} \propto (T-T_c)^n$ where n is ~1.2 [33] or ~1.5 [34]. This can be contrasted with the HTSC where a much stronger dependence on the temperature is found [35-37].

Initial measurements of the microwave resistance on separated $MgB_2$ grains (3 GHz) revealed a relatively high resistance below $T_c$ that could not be explained by a isotropic s-wave superconducting order parameter [38]. A conduction electron spin resonance study in the frequency range from 9 GHz to 225 GHz also reported an excess signal below $T_c$ [26]. A similar observation was made from an optical study on $MgB_2$ thin films with energies ranging from 0.5 meV to 3.7 meV [39] where the complex conductivity, $\tilde{\sigma} = \sigma_1 - i\sigma_2$, was reported. It was found that $\sigma_1$ below $T_c$ was larger than expected from the BCS model and $\sigma_2$ could not be fitted to a isotropic s-wave superconducting order parameter. This has implications for the penetration depth, $\lambda$, because $\sigma_2$ is related to $\lambda$ by $\lambda = (\mu_0 \omega \sigma_2)^{-1/2}$, where $\mu_0$ is the vacuum permeability and $\omega$ is the angular frequency. The observation that $\lambda$ could not be fitted to the isotropic s-wave BCS model was also reported in a muon spin rotation study where it was suggested that there exist nodes in the superconducting gap [21]. An optical study in the energy range from 4 meV to 14 meV concluded that $2\Delta/k_BT$ was significantly below that expected in the weak coupling BCS model [14]. It was found that $\sigma_2$ could be fitted to an isotropic s-wave superconducting gap when $\sigma_2$ at 3 meV was plotted against temperature. However, this $\sigma_2$ yields a very high value of the penetration depth (300 nm), much larger than that found by the muon spin rotation study (85 nm [21]) and greater than that determined from the magnetization data (140 nm [30], 180 nm [40]). A higher frequency far infrared study provided evidence that $MgB_2$ is in the dirty limit [41].

It is clear that further studies are required to resolve the existing controversies regarding the properties of the new superconductor, $MgB_2$. Some of the properties may vary due to the sample preparation conditions so that reports of resistivity, penetration depth, $H_{c2}$ and $H_{irr}$, need not give the same values. However, the use of different experimental techniques on the same well-characterized sample should provide the consistency required to help to resolve some of the puzzling issues. In this paper we report the results from magnetization measurements and microwave surface impedance measurements on a $MgB_2$ sample sintered at high temperature and high pressure. This sample has been characterized previously in a SEM study and its dc resistivity and Hall coefficient were reported [3,42,43].

**Experimental Details**

The MgB$_2$ sample was prepared by high pressure (3 GPa) synthesis using commercial MgB$_2$ powder as described elsewhere [3,42,43]. The sample was sintered at 950 °C and then quenched to room temperature. The density of the sample produced by this preparation method was 2.48 gcm$^{-3}$. Scanning electron microscopy (SEM) shows that the grains are well-connected and it is not possible to distinguish differences between individual grains within the resolution of the TEM [41].

Magnetization measurements were made using a SQUID magnetometer and applied magnetic fields of up to 60 kGauss. Ac magnetization measurements were made with an ac magnetic field of 0.05 Gauss and a frequency of 1 kHz. The dc magnetization data shows a small ferromagnetic component that is temperature-independent for temperatures less than 300 K. This is likely to be due to the Fe impurity clusters in this sample. Similar observations were reported by other researchers [21,28,44]. We estimate the Fe clusters fraction to be no more that 0.02 % of the total sample mass from the saturation magnetization of the ferromagnetic impurity.

Microwave measurements at 9.3 GHz were made using the intracavity method. The sample was mounted on a sapphire cold finger and placed in the center of the elliptic $_e$TE$_{111}$ cavity where the microwave electric field is maximum. The cavity was held at liquid helium temperature, and the heater and sensor assembly, mounted on the sapphire holder, enabled the sample temperature to be varied. The dc magnetic field was perpendicular to the microwave electric field. The measured quantities were the Q-factor and the resonant frequency of the cavity loaded with the sample. The cavity is made of copper with the walls coated with silver and a thin layer of gold in order to protect the walls from oxidation. This results in an unloaded Q-factor of $2 \times 10^4$, which is lower than the value attainable with superconducting cavities, but offers the advantage that measurements can be made in an applied magnetic field. With the recently introduced modulation technique [45], we were able to measure the change of the Q-factor with a relative precision of $10^{-3}$. The resonant frequency was monitored by a microwave counter. The complex frequency shift is given by [46],

$$\frac{\Delta \tilde{\omega}}{\omega} = \frac{\Delta f}{f} + i\Delta(\frac{1}{2Q}) = ia\tilde{Z}_s = a(-X_s + iR_s), \tag{1}$$

where $a$ contains the geometric parameters of the sample and cavity. The surface impedance is related to the complex conductivity by the expression $\tilde{Z}_s = \sqrt{i\mu_0 \omega / \tilde{\sigma}}$. The complex frequency shift in equation (1) is defined relative to that from an idealized perfect conductor with the same geometry as the actual sample. Since a perfect conductor does not give rise to microwave absorption, we take the empty cavity Q-factor as the reference point from which $\Delta(1/2Q)$ is monitored. The frequency offset is determined by the requirement that in the normal state $R_{sn}=X_{sn}$. It is convenient to normalize all the measured quantities to their values at some temperature in the normal state. The geometric factor $\boldsymbol{a}$ is then eliminated and the normalized measured quantities are related to the normalized real $\sigma_1/\sigma_n$ and imaginary $\sigma_2/\sigma_n$ components of the complex conductivity.

**Results and Analysis**

In can be seen from ac susceptibility measurements in figure 1a that the superconducting transition at 38.3 K is very sharp and the superconducting transition width is less than 1 K. This attests to the good quality and connectivity of the MgB$_2$ sample. It should be noted that measurements on commercial MgB$_2$ powder and MgB$_2$ sintered at one atmosphere can result in broad transitions into the Meissner state [31,44]. The corresponding dc M/H is plotted in figure 1b for an applied magnetic field of -25 Gauss in both the zero-field-cooled (solid curve) and field-cooled (dashed curve) conditions. The difference between the zero-field-cooled and field-cooled curves is due to trapped magnetic flux. Note that the remnant magnetic field from the SQUID superconducting magnet is ~6 Gauss and hence there is some trapped magnetic flux even in the nominally zero-field-cooled condition.

Two important parameters for characterizing type II superconductors are $H_{irr}$ and $H_{c2}$. The magnetic irreversibility is apparent in figure 2a where we show the difference between the field-cooled and zero-field-cooled dc magnetization for different applied magnetic fields. The irreversibility temperature for a given field, $T_{irr}(H)$, is defined by the departure from zero of the corresponding curves plotted in figure 2a. It can be seen that $T_{irr}(H)$ decreases with increasing magnetic field. These data are plotted in figure 2b as the inverted, $H_{irr}(T)$ (open circles). We show by the dashed curve in figure 2b that $H_{irr}$ can be fitted to $H_{irr}(T)=H_{irr}(0)(1-T/T_c)^n$ over the experimental temperature range where we find that n=1.4. The exponent is comparable to that found when a similar function is fitted to the $H_{irr}$ data of Bugoslavsky *et al.* (~1.5) [34] and it is greater than that obtained from the data of Fuchs *et al.* (~1.2) [33], where $H_{irr}$ was determined from the field-swept magnetization data. However, it is significantly greater than that observed by Larbalestier *et al.* (1.0) [31] where $H_{irr}$ was obtained by extrapolating the Kramer curves. It is also greater than that reported by Kim *et al.* (1.0) [32] where $H_{irr}$ was defined as the field where the initial negative maximum in the magnetization is observed when the magnetization is measured as a function of increasing magnetic field. However, it is less than that observed in the HTSC near $T_c$ (1.5 to 2.0) [35-37].

The origin of the magnetic irreversibility in the HTSC is still not resolved and the interpretations include, flux-line pinning or flux creep, 3D flux-lattice melting and a fluctuation-induced 3D-2D flux-line transition. The third mechanism is unlikely to be valid in $MgB_2$ because the anisotropy in the penetration depth and superconducting coherence length is believed to be small. We note that the exponent in our $MgB_2$ sample is close to that obtained from zero-field-cooled and field-cooled magnetization measurements on $Nb_3Sn$ (1.37 [47]) where the irreversibility was fitted to the flux-lattice melting model of Houghton, Pelcovitis and Sudbø [48]. It is therefore possible that the irreversibility in our $MgB_2$ sample is also due to flux-lattice melting. However, as we have mentioned in relation to the HTSC, it is not possible on the basis of the critical exponent to uniquely determine the origin of $H_{irr}$ [37]. We show later that the magnetic irreversibility in the microwave region is associated with vortex depinning.

It can be seen in figure 2b that $H_{c2}$ (solid circles) increases linearly with decreasing temperature. $H_{c2}$ was obtained from the temperature-dependent magnetization data where the temperature at which the magnetization begins to decrease in an applied magnetic field defines $H_{c2}(T)$. We find that $dH_{c2}/dT=5100$ Gauss/K near $T_c$ which is comparable to that found by Larbalestier *et al.* (~5000 Gauss/K) [31] and slightly greater than that found by Finnemore *et al.* (4400 Gauss/K) [30]. However, other studies have found that $H_{c2}=H_{c2}*(1-T/T_c)^{1+\alpha}$ where $\alpha$ can be 0.25 or greater [23,27-29]. It has been stated that a positive curvature is also observed in the rare-earth nickel borocarbides, where it has been attributed to an effective two-band model for superconductors in the clean limit [27]. However, as mentioned earlier, a far infrared study provided evidence that $MgB_2$ is in the dirty limit [41]. The discrepancy between different studies may arise from different sample preparation conditions. Alternatively, it may be due to the experimental definition of $H_{c2}$ rather than underlying band structure effects. We shall deal with this problem in more detail later in the paper.

The temperature dependences of $R_s/R_{sn}$ and $X_s/X_{sn}$ are plotted in figure 3a. The temperature dependence of the surface resistance is similar to those reported earlier from microwave measurements [38,49,50]. In particular, there is an excess surface resistance below $T_c$ that can not be explained by a single s-wave superconducting gap, as we show later. We have extended previous studies by also measuring surface reactance. It can be seen that the transitions into the superconducting state are sharp and there is no evidence of the superconducting fluctuation effects that are observed in the HTSC. In the case of the HTSC, superconducting fluctuations are important near $T_c$ because of the very short c-axis coherence length (~0.1 nm in $Bi_2Sr_2Ca_2Cu_3O_{10+\delta}$ [51]). This can be contrasted with $MgB_2$ where the c-axis coherence length, estimated from $H_{c2}$, is equal to or greater than 2 nm [52].

Below $T_c$, the surface reactance shows a peak followed by a gradual decrease and saturation at low temperatures. The $R_s$ and $X_s$ data plotted in figure 3a do not appear unusual but when inverted to obtain the complex conductivity we observe a surprising result, as can be seen in figure 3b where we plot the temperature dependences of $\sigma_1/\sigma_n$ and $\sigma_2/\sigma_n$. These curves deviate significantly from those expected for an s-wave superconducting order parameter and within the weak-coupling BCS model (solid curves), where we use $2\Delta(0)=3.53 k_B T_c$. The huge increase of $\sigma_1/\sigma_n$ below $T_c$ is a consequence of the large $R_s/X_s$ ratio. Similar observations were reported in YBCO single crystals and interpreted as an unusual increase in the electron scattering time [53,54] with decreasing temperature. In the present case, such an interpretation can be ruled out since microwave resistance measurements at magnetic fields greater than $H_{c2}$ reveal no dramatic changes in the normal-state conductivity when compared with the normal-state conductivity at $T_c(H=0)$. One could consider impurities as the cause of this anomalous behavior. As mentioned earlier, dilute Fe impurities are present in this sample. However, the resistivity of the present sample is comparable to, or less than, those found in other sintered ceramic samples [29,55]. Furthermore, the gradual decrease of $R_s$ at low temperatures is not easy to interpret in terms of impurities. Since the sample is not a single crystal, it is possible that weak links could be the origin of the slowly decreasing $R_s$ at low temperatures. However, as we show later, magnetic field measurements reveal no sign of weak links in our sample. Finally, we consider the multiple gap structure observed in this compound. It is conceivable that such a structure is at the origin of the unusual behavior of $\sigma_1/\sigma_n$. If the gaps arise from the σ and π bands then it is possible that the temperature dependence of the two condensate densities are different and the resulting surface impedance problem could not be described by the simple expression used here. This interpretation has already been suggested by Zhukov *et al.* [38].

Looking at the temperature dependence of $\sigma_2/\sigma_n$, plotted in figure 3b, it is apparent that there is an unusually slow increase in $\sigma_2/\sigma_n$ with decreasing temperature just below $T_c$ and an almost linear increase with decreasing temperature in the low temperature region. This anomalous behavior of $\sigma_2/\sigma_n$ can not be due to impurities, since $\sigma_2$ does not depend on the scattering rate, but may arise from the multiple superconducting gaps. Our results extend the observed features of the complex conductivity in the submillimeter-wave region [39] to significantly lower frequencies. Our frequency is an order of magnitude lower than the lowest one used by Pronin *et al.* [39] and we find that the maximum in $\sigma_1/\sigma_n$ is far more pronounced and shifted to lower temperatures. Similarly, the temperature dependence of $\sigma_2/\sigma_n$ in our case is more anomalous than in the submillimeter-wave range. A linear extrapolation of $\sigma_2/\sigma_n$ in figure 3b would yield a ratio of about 100. Taking the value $\sigma_n = 5 \times 10^6 \, \Omega^{-1} m^{-1}$ at $T_c$ from the dc resistivity measurements on this sample [42] we find that $\sigma_2(0) = 5 \times 10^8 \, \Omega^{-1} m^{-1}$ and the zero temperature penetration depth is $\lambda_L = 160$ nm. This value is comparable with the one estimated from the magnetization data [30,40].

The temperature dependences of $R_s/R_{sn}$ and $X_s/X_{sn}$ are plotted in figure 4 for magnetic fields of 40 kGauss (open up triangles) and 80 kGauss (open down triangles) together with the data in zero field (filled circles). It can be seen that the normalized surface resistance is extended linearly from the normal state to lower temperatures until the transition into the superconducting state occurs. This indicates that there are no magnetoresistance effects observed in our sample in the normal state. A similar linear extension in the normal-state is also apparent in the surface reactance for temperature less than $T_c(H=0)$.

The magnetic and electronic properties of our sample are more apparent in figure 5 where we plot $R_s/R_{sn}$ and $X_s/X_{sn}$ against magnetic field and at different temperatures. It is clear that there are no weak links in our sample because weak links are known to cause a rapid initial increase in $R_s$ and $X_s$ especially in the sintered ceramic HTSC samples [56-58], which is not seen in figure 5. The absence of weak links indicates that the magnetic and electronic properties of our sample, measured at low field, are intrinsic. It also shows the technological potential of $MgB_2$ when compared with the HTSC because the electronic and magnetic properties of the main HTSC being developed for wire application, $Bi_2Sr_2Ca_2Cu_3O_{10+\delta}$, are significantly affected by weak links.

We estimate $H_{c2}$ from the curves in figure 5 by performing a linear extrapolation above $T_c$. This procedure is clear in figure 6 where we plot $R_s/R_{sn}$ at 33 K. The departure from the straight line, indicated by the arrow, is the temperature where $H_{c2}$ is equal to the applied magnetic field. The results obtained at various temperatures are summarized in figure 2b (filled up triangles). It is apparent that this method of determining $H_{c2}$ yields similar results as the method based on the magnetization data.

It is also possible to estimate $H_{c2}$ from the $R_s$ and $X_s$ data plotted in figure 5 by considering the effective conductivity in the mixed state which can be expressed as [59],

$$\frac{1}{\tilde{\sigma}_{eff}} = \frac{1-(H/H_{c2})(1-i\omega_0/\omega)^{-1}}{(1-H/H_{c2})(\sigma_1-i\sigma_2)+\sigma_n H/H_{c2}} + \frac{H}{H_{c2}\sigma_n}\frac{1}{1-i\omega_0/\omega} \qquad (2)$$

where $\omega$ is the microwave angular frequency and $\omega_0$ is the depinning angular frequency. It can be seen that $\tilde{\sigma}_{eff}$ reduces to the zero-field conductivity, $\sigma_1-i\sigma_2$, when H is zero. Furthermore, as $H \rightarrow H_{c2}$ the effective conductivity tends towards the normal state conductivity, $\sigma_n$. However, in the mixed state $\tilde{\sigma}_{eff}$ is field dependent. The ratio $H/H_{c2}$ defines the volume fraction of the sample taken by the vortex cores. The concept of an effective conductivity is valid when the distance between vortices is much smaller than the penetration depth. Hence, one should not apply it at low fields. This model assumes that the radius of the vortex cores is $\xi = \sqrt{\Phi_0/2\pi H_{c2}}$, where $\Phi_0$ is the flux quantum. At a constant temperature, the density of vortices increases with increasing magnetic field but all the vortices have the same fixed radius, $\xi$. Thus, $H_{c2}$ in equation 2 is not only the field above which the normal state is reached, it is also a constant in the ratio $H/H_{c2}$ for all magnetic fields less than $H_{c2}$. The depinning angular frequency parameter, $\omega_0$, depends on the pinning potential experienced by the vortices driven by the oscillating microwave current [60].

We plot is figure 7, $H_{c2}$ and $\omega_0/\omega$ against applied magnetic field where $H_{c2}$ and $\omega_0/\omega$ were obtained using equation 2, the data in figure 5 and the zero-field values of $\sigma_1/\sigma_n$ and $\sigma_2/\sigma_n$ plotted in figure 3b. The effective conductivity was obtained from, $1/\tilde{\sigma}_{eff} = \tilde{Z}_s^2/i\mu_0\omega$, using the data plotted in figure 5. We first discuss the depinning frequency. At a fixed temperature, $\omega_0$ decreases with increasing applied magnetic field, thus showing that the pinning potential is reduced as the vortex density increases. This observation indicates that collective pinning of vortices occurs over the whole magnetic field region covered in figure 7. In contrast, a field independent $\omega_0$ is observed in HTSC at fields much lower than $H_{c2}$, which is due to individual vortex pinning [60]. When the pinning is individual the pinning potential depends only on temperature and $\omega_0$ is independent of magnetic field at a given temperature. In collective pinning, however, the effective pinning potential depends on the interactions between vortices and this brings about the field dependence of $\omega_0$. We note that the vortex solid occurs in the irreversible region where $w_0/w \gg 1$, while for $w_0/w < 1$ there is a gradual transition into the reversible region with decreasing $w_0/w$. The flux-flow regime occurs when $w_0/w \ll 1$. Therefore, the data in figure 7b indicate that the magnetic irreversibility observed in the microwave region is associated with vortex depinning. However, the irreversibility field obtained in the microwave region is less than that observed in the dc case. This can be seen in figure 2b where we plot the magnetic field where $w_0/w = 1$. This is expected since depinning at microwave frequencies is a collective excitation of the flux line lattice but with flux still trapped in the sample. Depinning in the context of dc magnetization means that the flux has to move across the sample surface. We expect that these different processes appear at different energies. For the dc magnetization case, depinning is thermally activated and only controlled by the temperature. In the microwave case, depinning is manifested when the viscous drag force, which is proportional to the velocity of the oscillating flux lines, becomes comparable to the restoring force due to pinning in a potential well. This occurs at a temperature lower than that required for the dc magnetization depinning. We show by the dotted curves in figure 2b that the magnetic field where $w_0/w = 1$ is proportional to $(1-T/T_c)^n$ where n=1.25. This line represents the dynamic irreversibility at 9.3 GHz.

It is apparent in figure 7a that $H_{c2}$, calculated using the mixed state model discussed above, is more complex that those plotted in figure 2b. It particular, the inversion process reveals $H_{c2}$ values that increase with increasing applied magnetic field. In an single crystal superconductor one should get a single constant value for $H_{c2}$ since it reflects the value of $x$, which is constant for a given temperature. We attribute the increase in $H_{c2}$ with increasing field to an intrinsic anisotropy in $H_{c2}$ and random orientations of the grains. When the applied magnetic field is increased above the lowest value of the anisotropic $H_{c2}$, some of the grains are brought to the normal state, and only those grains with orientations yielding higher $H_{c2}$ remain superconducting. In this way, the increasing applied magnetic field gradually changes the average $H_{c2}$ of the grains that still remain in the superconducting state. We note that Eq. (2) is highly nonlinear and the value of $H_{c2}$, calculated by the inversion procedure from the experimental data, is not a simple average of the intrinsic $H_{c2}$ from the individual grains. The result in figure 7 merely proves that there is no single value for $H_{c2}$ at a given temperature, which indicates that the sample has randomly oriented grains with an intrinsic anisotropy in $H_{c2}$. We note that the value of $H_{c2}$, obtained from the dc magnetization data and the temperature where the microwave resistance begins to decrease, is the magnetic field where the last grains in the sample are brought into the normal state. Thus, for an anisotropic and unoriented superconductor, this method will provide $H_{c2}$ values that are close to the highest intrinsic upper critical field $H_{c2}^c$, which is the one obtained for the magnetic field applied along the c-axis.

**Conclusion**

In conclusion, we have performed magnetization and microwave studies on $MgB_2$. We find an unusually large microwave surface resistance below $T_c$ and $\sigma_1$ and $\sigma_2$ do not follow the curves expected for a s-wave BCS superconductor. We speculate that this may be due to the two superconducting gaps reported in other studies. The magnetic field dependent microwave impedance shows no evidence of Josephson weak-links which indicates that the crystal defects and intergranular regions are significantly smaller that the superconducting coherence length. Furthermore, we find that the upper critical field is linearly dependent on temperature and the dc irreversibility field coefficient is larger than that found in some of the previous studies. We show from our analysis of the microwave data that the vortex pinning is collective, which is different from that observed in the HTSC where individual vortex pinning is observed. Furthermore, the magnetic irreversibility in the microwave region is due to dynamic vortex depinning. We also find that the magnetic field dependent microwave impedance can not be fitted to a model with one $H_{c2}$. This observation can be interpreted in terms of an intrinsic anisotropy in $H_{c2}$ and randomly oriented grains in the sample.

**Acknowledgements**

We acknowledge funding support from the Croatian Ministry of Science and Technology (AD, DP and MP), the New Zealand Marsden Fund (GVMW), the Alexander von Humboldt Foundation (GVMW) and the Ministry of Science and Technology of Korea through the Creative Research Initiative Program (C. U. Jung, Min-Seok Park and Sung-Ik Lee.).

**Figures**

**Figure 1:** (a) Plot of the real part of the ac susceptibility against temperature. (b) Plot of the zero-field-cooled (solid curve) and field-cooled (dashed curve) magnetization divided by the field for an applied magnetic field of –25 Gauss.

**Figure 2:** (a) Plot of the magnetization difference between field-cooled and zero-field-cooled measurements for applied magnetic fields of 60 kGauss, 50 kGauss, 40 kGauss, 30 kGauss, 20 kGauss, 10 kGauss and 25 Gauss. The arrow indicated increasing applied magnetic field. (b) Plot of $H_{c2}$ (filled circles) and $H_{irr}$ (open circles) against temperature. The curves are fits to the data obtained from the magnetization measurements as described in the text. Also shown is $H_{c2}$ (filled up triangles) determined from the microwave $R_s/R_{sn}$ using the linear extrapolation method described in the text and the magnetic field where $w_0/w$, plotted in figure 7b, is equal to 1 (open up triangles). The dotted curve is a fit to the microwave data as described in the text.

**Figure 3:** (a) Plot of $R_s/R_{sn}$ (filled circles) and $X_s/X_{sn}$ (open up triangles) against temperature. (b) The corresponding $\sigma_1/\sigma_n$ (filled circles) and $\sigma_2/\sigma_n$ (open up triangles) plotted against temperature. The data are normalized to the values at 40 K. The solid lines show the calculated BCS curves in the weak coupling limit.

**Figure 4:** (a) Plot of $R_s/R_{sn}$ against temperature for applied magnetic fields of 0 Gauss (filled circles), 40 kGauss (open up triangles) and 80 kGauss (open down triangles). (b) Plot of $X_s/X_{sn}$ against temperature for applied magnetic fields of 0 Gauss (filled circles), 40 kGauss (open up triangles) and 80 kGauss (open down triangles).

**Figure 5:** (a) Plot of $R_s/R_{sn}$ against magnetic field for temperatures of 5, 10, 15, 20, 25, 30, 33, 35 and 37 K. (b) Plot of $X_s/X_{sn}$ against magnetic field for the same temperatures. The arrows indicate increasing temperature.

**Figure 6:** Plot of $R_s/R_{sn}$ against magnetic field at 33 K on an expanded scale. The arrow indicates the field where $R_s/R_{sn}$ starts to deviate from a straight line, which is taken as $H_{c2}$.

**Figure 7:** Plot of $H_{c2}$ and $w_0/w$ obtained by numerically inverting of the data from figure 5 as described in the text. The arrows indicate increasing temperatures from 5 K to 33 K.

Figure 1
Dulcic *et al.*
Phys. Rev. B

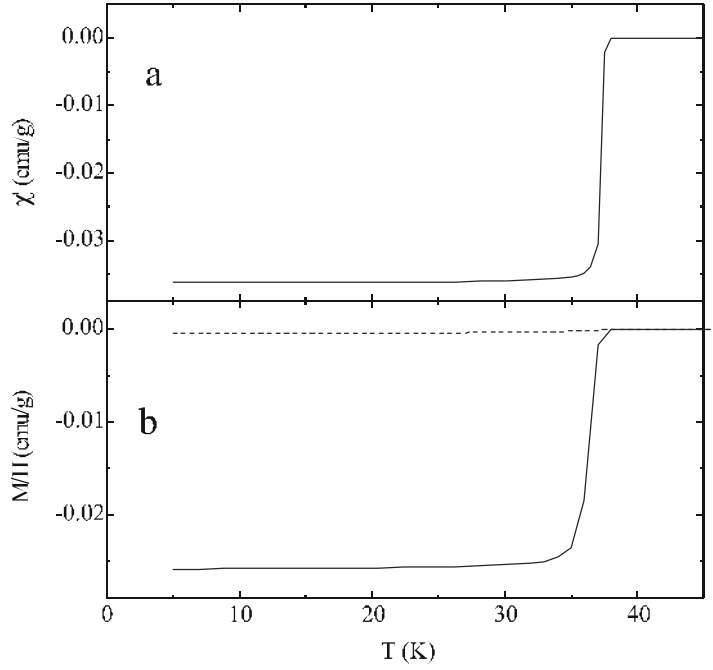

Figure 2
Dulcic *et al.*
Phys. Rev. B

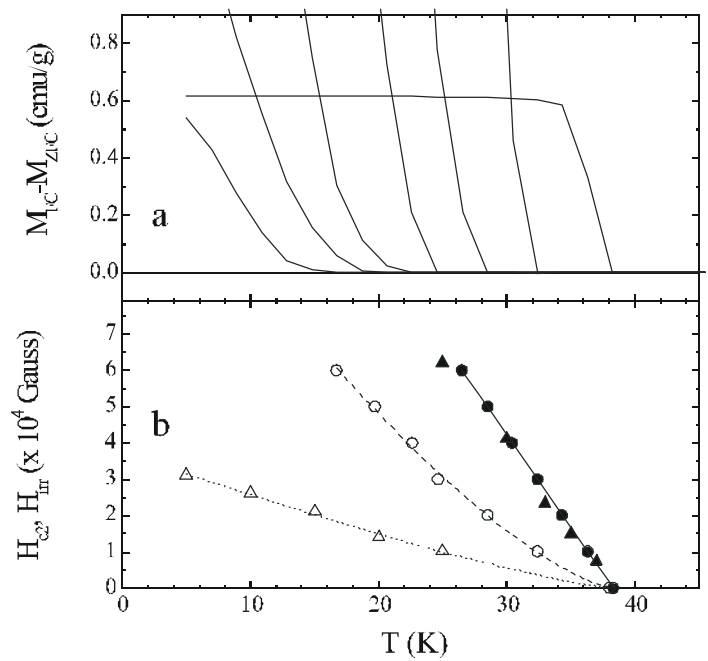

Figure 3
Dulcic *et al.*
Phys. Rev. B

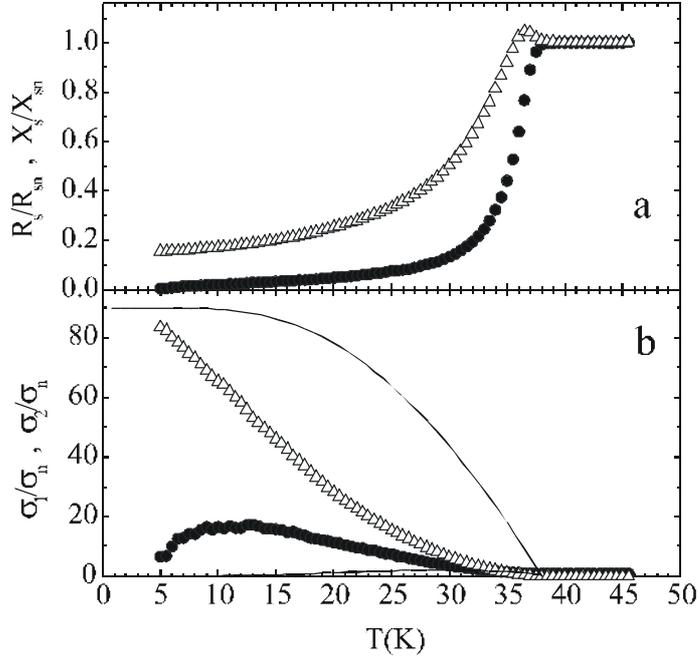

Figure 4
Dulcic *et al.*
Phys. Rev. B

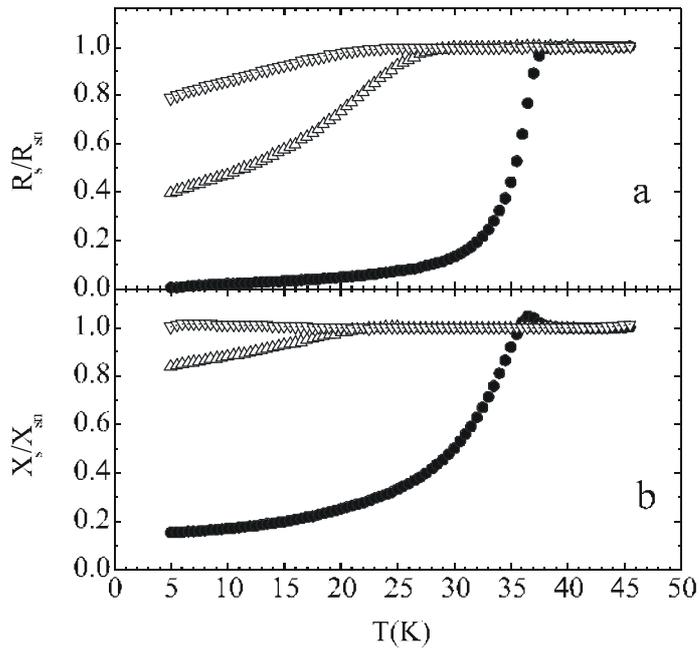

Figure 5
Dulcic *et al.*
Phys. Rev. B

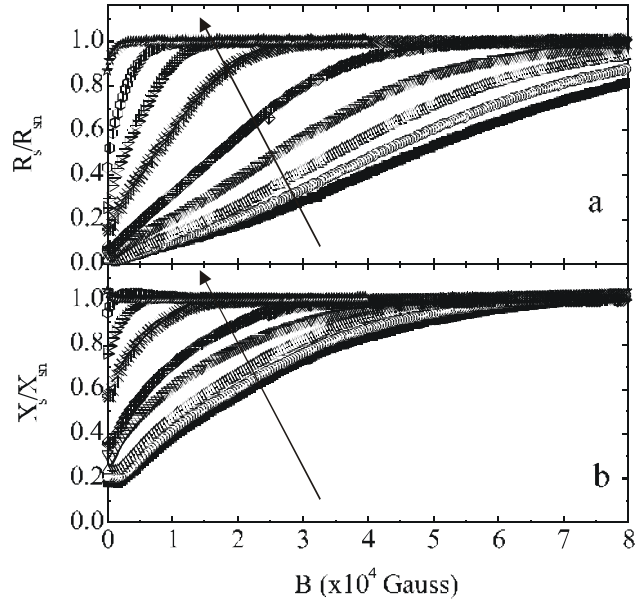

Figure 6
Dulcic *et al.*
Phys. Rev. B

Figure 7
Dulcic *et al.*
Phys. Rev. B

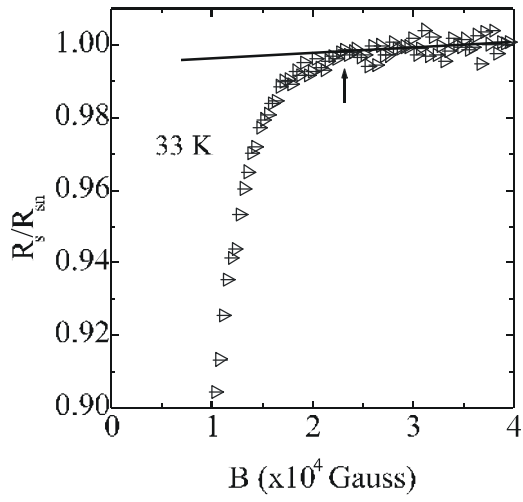
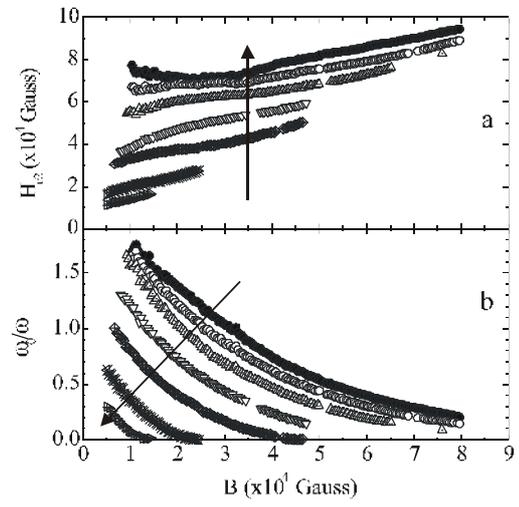